%% file: main.tex
\documentclass[superscriptaddress,preprintnumbers,amsmath,amssymb, aps, prl, twocolumn, nofootinbib]{revtex4-2}
\usepackage[utf8]{inputenc}
\usepackage{graphicx}
\usepackage{latexsym,amsmath,amssymb,amsthm,lmodern,float,url,bm}
\usepackage{natbib}
\usepackage{color}
\usepackage{microtype}
\usepackage{import}
\usepackage{bbold}
\usepackage[plain]{fancyref}
\usepackage{varioref}
\usepackage{slashed}
\usepackage{multirow}
\usepackage{physics}
\usepackage{comment}
\usepackage{slashed}
\allowdisplaybreaks

\usepackage[colorlinks=true,backref=false, linktocpage=true,
citecolor=blue,urlcolor=blue,linkcolor=blue,pdfpagemode=UseOutlines]{hyperref}
\hypersetup{%
  bookmarksnumbered=true,
  pdftitle = {},
  pdfsubject = {},
  pdfauthor = {},
  pdfkeywords = {}
}

\let\Im\undefined

\DeclareMathOperator{\Im}{Im}

\DeclareMathOperator{\im}{Im}
\newcommand{\BDPi}{$B\rightarrow D \pi \ell \nu$ }
\newcommand{\BDeta}{$B\rightarrow D \eta \ell \nu$ }

\definecolor{darkgreen}{rgb}{0.0, 0.2, 0.13}
\definecolor{darkspringgreen}{rgb}{0.09, 0.45, 0.27}

\begin{document}
\preprint{FERMILAB-PUB-23-662-T\quad ZU-TH 67/23}
\title{A model-independent description of \BDPi decays}

\author{Erik J. Gustafson}
\affiliation{Fermi National Accelerator Laboratory, Batavia,  Illinois, 60510, USA}

\author{Florian Herren}
\affiliation{Fermi National Accelerator Laboratory, Batavia,  Illinois, 60510, USA}
\affiliation{Physics Institute, Universität Zürich,
Winterthurerstrasse 190, CH-8057 Zürich, Switzerland}

\author{Ruth S. Van de Water}
\affiliation{Fermi National Accelerator Laboratory, Batavia,  Illinois, 60510, USA}

\author{Raynette van Tonder}
\affiliation{Department of Physics, McGill University, 3600 rue University, Montréal, Québec, H3A 2T8, Canada}
\affiliation{Institut für Experimentelle Teilchenphysik, Karlsruhe Institute of Technology (KIT), D-76131 Karlsruhe, Germany}

\author{Michael L. Wagman}
\affiliation{Fermi National Accelerator Laboratory, Batavia,  Illinois, 60510, USA}

\date{\today}

\begin{abstract}
We introduce a new parameterization of \BDPi form factors using a partial-wave expansion and derive bounds on the series coefficients using analyticity and unitarity.  This is the first generalization of the model-independent formalism developed by Boyd, Grinstein, and Lebed for $B \to D \ell \nu$ to semileptonic decays with multi-hadron final states, and enables data-driven form-factor determinations with robust, systematically-improvable uncertainties. 
Using this formalism, we extract the form-factor parameters for $B \to D_2^\ast(\to D\pi) \ell \nu$ decays in a model-independent way from fits of data from the Belle Experiment. We find that the semileptonic data are compatible with the presence of two poles in the $D\pi$ S-wave channel, which is the scenario preferred by nonleptonic decays and unitarized chiral perturbation theory.
\end{abstract}

\maketitle

\textbf{\textit{Motivation --- }}
Experimental measurements of tree-level semileptonic $B$-meson decays enable theoretically clean determinations of the Cabibbo-Kobayashi-Maskawa (CKM) matrix elements $|V_{ub}|$ and $|V_{cb}|$, allowing for sensitive tests of the Standard Model by overconstraining the CKM unitarity triangle~\cite{Charles:2004jd,HFLAV:2019otj,UTfit:2022hsi}. Further, $|V_{ub}|$ and $|V_{cb}|$ are parametric inputs to predictions for loop-level flavor-changing processes that are sensitive to new high-scale physics beyond the reach directly detectable by the LHC~\cite{DiCanto:2022icc,Artuso:2022ouk}.

A major challenge for both inclusive and exclusive determinations of $|V_{ub}|$ is suppressing the CKM-favored $B \rightarrow X_{c} \ell \nu$ background, which exhibits a similar experimental signature and is $\mathcal{O}(100)$ times more abundant than $B \rightarrow X_{u} \ell \nu$ decays. The background subtraction process is further complicated by the orbitally excited states, collectively referred to as $D^{**}$, whose kinematic distributions remain poorly understood and branching fractions exhibit uncertainties of approximately 20\% \cite{ParticleDataGroup:2022pth}. In measurements performed by the Belle and Belle II Collaborations, the remaining “gap” between the sum of all considered exclusive modes and the inclusive $B \rightarrow X \ell \nu$ branching fraction, comprising unmeasured non-resonant $B \rightarrow X_{c} \ell \nu$ decays, is generally treated in simulation by assuming a composition of equal parts of $B \rightarrow D^{(*)} \eta \ell \nu $ decays, as prescribed in Ref.~\cite{Belle:2021idw}. Because neither experimental evidence nor theoretical predictions exist for $B \rightarrow D^{(*)} \eta \ell \nu $ decays, a 100\% uncertainty is assumed for the corresponding branching fractions. For these reasons, the $X_{c} \ell \nu$ modeling uncertainty is hard to quantify and becomes dominant for studies of inclusive $B \rightarrow X_{c/u} \ell \nu$ decays \cite{Belle:2021idw,Belle:2006jtu,Belle:2006kgy,Belle:2021eni,Belle-II:2022evt}.

Exclusive measurements relying on tagged methods, in which machine learning algorithms are employed to fully reconstruct the companion $B$ meson through exclusive decay modes \cite{Feindt:2011mr, Keck:2018lcd}, do not rely as directly on $X_{c} \ell \nu$ modeling as inclusive analyses. However, significant differences in these reconstruction algorithms' performances between data and simulation is accounted for by performing a calibration using a decay with a well known branching fraction: inclusive $B \rightarrow X \ell \nu$ \cite{Belle-II:2020fst}. This calibration, in turn, becomes a leading source of systematic error for tagged analyses \cite{Belle-II:2022pir,BelleII:2023lst,Belle:2023bwv}. In addition, the limited knowledge of $B\rightarrow D^{\ast\ast}\ell/\tau\nu$ branching fractions and form factors are large systematic uncertainties in studies of rare processes or lepton flavor universality tests such as $B \rightarrow K \nu \nu$ and $R(X_{\tau/\ell})$ at Belle II \cite{Belle-II:2023esi,Belle-II:2023aih} or $R(D^\ast)$ at the LHCb experiment \cite{LHCb:2023uiv,LHCb:2023zxo}.

The most commonly used description of $B \rightarrow D^{\ast\ast} \ell \nu$ decays is the Leibovich-Ligeti-Stewart-Wise (LLSW) parameterization \cite{Leibovich:1997tu,Leibovich:1997em}, extended to include $\mathcal{O}(\alpha_s)$ corrections and relaxing several assumptions with central values from the fit given in~Refs. \cite{Bernlochner:2016bci,Bernlochner:2017jxt}. This parameterization includes a single $D_0^\ast$ resonance. Studies in the context of unitarized chiral perturbation theory (UChPT), however, have shown that the scalar member of the $D^{\ast\ast}$ family, the $D_0^\ast(2300)$, is an overlap of two states with poles near $(2.1 - i 0.1)$ and $(2.45 - i 0.13)$ GeV~\cite{Albaladejo:2016lbb,Du:2017zvv,Guo:2017jvc}. Consequently, the S-wave lineshape is not described by a simple Breit-Wigner distribution, but has a more complex structure.
This conclusion is supported by lattice quantum chromodynamics (LQCD) calculations of isospin-1/2 $D\pi$ scattering~\cite{Moir:2016srx,Gayer:2021xzv,Yan:2023gvq} and a reinterpretation~\cite{Du:2020pui} of the partial-wave analysis of $B^+ \rightarrow D^- \pi^+ \pi^+$ decays by LHCb~\cite{LHCb:2016lxy}. 
Further, in Ref.~\cite{LeYaouanc:2022dmc}, Le Yaouanc, Leroy, and Roudeau point out that in the fits to the LLSW parametrization tail effects from the $D^\ast$ resonance are omitted and consequently overestimates the $D_0^\ast$ contribution.

To address these and other limitations of existing parameterizations,
in this letter we develop the first model-independent description of resonant and non-resonant \BDPi decays based on analyticity and unitarity. We then apply our formalism, which accommodates arbitrary lineshapes, to fit experimental spectrum measurements and draw conclusions about the pole structure of the S-wave channel.

\textbf{\textit{Form-factor parameterization --- }}
Semileptonic \BDPi decays are characterized by five kinematic variables: the momentum transfer square $q^2$,\footnote{It is sometimes useful to instead consider the dependence on the recoil parameter $w = (M_B^2 + M_{D\pi}^2 - q^2)/(2M_B M_{D\pi})$.} the helicity angle of the charged lepton $\cos\theta_l$, the helicity angle of the $D$ meson $\cos\theta$, the azimuthal angle between the $\ell\nu$ and $D\pi$ planes $\chi$, and the invariant mass of the hadronic system $M_{D\pi}$.

Form-factor decompositions for charged-current semileptonic decays involving two final state hadrons have been performed for $B\rightarrow \pi\pi\ell\nu$ decays \cite{Faller:2013dwa,Kang:2013jaa} and involve a partial-wave decomposition in $\cos\theta$ to disentangle contributions from different hadronic resonances. Following a similar strategy, we express the \BDPi hadronic matrix elements as
\begin{widetext}
\begin{equation}
\begin{split}
    &\left\langle D(p_D)\pi(p_\pi)|V^\mu|B(p_B)\right\rangle = i\epsilon^\mu_{\nu\rho\sigma}p_{D\pi}^\rho p_B^\sigma \sum_{l > 0} L^{(l),\nu}g_l(q^2,M_{D\pi}^2)~,\\
    &\left\langle D(p_D)\pi(p_\pi)|A^\mu|B(p_B)\right\rangle =
    \frac{1}{2}\sum_{l > 0} \left(L^{(l),\mu}+\frac{4}{\lambda_B}\left[(p_B\cdot p_{D\pi})q^\mu - (p_{D\pi}\cdot q)p_B^\mu\right]L^{(l),\nu}q_\nu\right)f_l(q^2,M_{D\pi}^2)\\
    &+\frac{M_{D\pi}(M_B^2-M^2_{D\pi})}{\lambda_B}\left[(p_B + p_{D\pi})^\mu - \frac{M^2_B-M^2_{D\pi}}{q^2}q^\mu\right]\sum_{l> 0} L^{(l),\nu}q_\nu \mathcal{F}_{1,l}(q^2,M_{D\pi}^2) +M_{D\pi}\frac{q^\mu}{q^2} \sum_{l > 0} L^{(l),\nu}q_\nu \mathcal{F}_{2,l}(q^2,M_{D\pi}^2) \\
     &+ \left[(p_B + p_{D\pi})^\mu - \frac{M^2_B-M^2_{D\pi}}{q^2}q^\mu\right]f_+(q^2,M_{D\pi}^2) + \frac{M^2_B-M^2_{D\pi}}{q^2}q^\mu f_0(q^2,M_{D\pi}^2).\label{eq:ffs1}
\end{split}
\end{equation}
\end{widetext}
The vector $L^{(l)}$ is related to the angular momentum  of the final-state hadron system in the $B$-meson rest frame, and is uniquely defined via
\begin{equation}
\begin{split}
  L^{(l)}_\mu q^\mu  &= M_B W^l P_l(\cos\theta) \\
    L_\mu^{(l)}p^\mu_{D\pi} &= 0\,,
    \label{eq:Lvec}
    \end{split}
\end{equation}
where $W = |\vec{q}| |\vec{p_D}|/(M_B M_{D\pi})$ and $P_l$ are the Legendre polynomials. The threshold factor $\lambda_B = M_B^4 + M_{D\pi}^4 + q^4 - 2(M_B^2M_{D\pi}^2+M_{D\pi}^2q^2 + q^2 M_B^2)$.

The standard expressions for $B\rightarrow D^\ast\ell \nu$ and $B\rightarrow D_2^\ast \ell \nu$ decays~\cite{Wirbel:1985ji,Mandal:2019vwq} are recovered from the $l=1$ and $l=2$ terms by replacing $M_{D\pi}$ and $L^{(l)}$ with the corresponding masses and polarization vectors.

Using Eqs.~\eqref{eq:ffs1} and~(\ref{eq:Lvec}), it is straightforward to derive the \BDPi differential decay rate.
After performing the angular integration and dropping terms that are helicity suppressed, we obtain the double differential decay rate for massless leptons, 
\begin{widetext}
\begin{equation}
\frac{\mathrm{d}^2\Gamma}{\mathrm{d}M^2_{D\pi}\mathrm{d}q^2} = \frac{G_F^2|V_{cb}|^2}{(4\pi)^5}M_B  \left(W\frac{\lambda_B}{M_B^2}\frac{4|f_+|^2}{3}+ M_{D\pi}^2\sum_{l>0}\frac{4 W^{2l+1}}{3(2l+1)}\left[(M_B^2-M_{D\pi}^2)^2\frac{|\mathcal{F}_{1,l}|^2}{\lambda_B}+\frac{(l+1)}{l}q^2\left(|g_l|^2+\frac{|f_l|^2}{\lambda_B}\right)\right]\right)\,,
\end{equation}
\end{widetext}
which will be used later in our analysis.
The fully general five-fold differential decay rate allowing for interference effects between different partial waves is provided in the supplemental material~\cite{supplemental}.

\textbf{\textit{Unitarity bounds} --- }
Model-independent constraints on the \BDPi form factors arise from analyticity and unitarity. We begin with the two-point functions
\begin{equation}
  \Pi_{(J)}^{L/T}(q) \equiv i\int d^4x\  e^{i q\cdot x}\,  \left< 0\right| J^{L/T}(x) \, J^{L/T}(0) \left| 0\right>,\label{eq:2pt}
\end{equation}
where $J^\mu$ denotes a $b\rightarrow c$ flavor-changing vector or axial-vector current and $L/T$ denotes the component longitudinal or transverse to $q^\mu$.
Susceptibilities $\chi_{(J)}^{L/T}$ are defined from derivatives of $\Pi_{(J)}^{L/T}(q)$ as\vspace{-5pt}
\begin{equation}
    \begin{split}
    &\chi^L_{(J)}(Q^2) \equiv \frac{\partial \Pi_{(J)}^L}{\partial q^2}\Big|_{q^2=Q^2} = \frac{1}{\pi}\int_{0}^\infty\mathrm{d}q^2 \frac{\im \Pi_{(J)}^L(q^2)}{(q^2-Q^2)^2}~,\\
    &\chi^T_{(J)}(Q^2) \equiv \frac{1}{2}\frac{\partial^2 \Pi_{(J)}^T}{\partial (q^2)^2}\Big|_{q^2=Q^2} = \frac{1}{\pi}\int_{0}^\infty\mathrm{d}q^2 \frac{\im \Pi_{(J)}^T(q^2)}{(q^2-Q^2)^3}~,\label{eq:disp}
    \end{split}
\end{equation} 
and are related to integrals over the imaginary part of $\Pi_{(J)}^{L/T}(q)$ via dispersion relations.
The susceptibilities $\chi^{L/T}_{(V/A)}$ at $Q^2 = 0$ have been computed in perturbation theory to ${\mathcal O}(\alpha_S^2)$ in Ref.~\cite{Grigo:2012ji} and with nonperturbative lattice QCD in Refs.~\cite{Martinelli:2021frl,Melis:2024wpb}.
Separately, the optical theorem relates $\im \Pi_{(J)}^T(q^2)$ to a sum of squared amplitudes for all intermediate states that can appear between the currents in Eq.~\eqref{eq:2pt}.
This sum includes terms with the matrix element $\left< \overline{B} D \pi |J^{L/T}|0\right>$, which is related to $\left\langle D\pi |J^{L/T} |B\right\rangle$ by crossing symmetry and can therefore be parameterized by the form-factor decomposition in Eq.~\eqref{eq:ffs1}.
The contribution of the \BDPi channel to the dispersion relations in Eq.~\eqref{eq:disp} is then given by evaluating the phase space integrals arising in the sum over states,
\begin{widetext}
\begin{equation}
    \begin{split}
        \Im \Pi_{A}^L\Big|_{D\pi} &= \frac{1}{64\pi^3}\frac{M_B^4}{q^4} \int_{(M_D + m_\pi)^2}^{(\sqrt{q^2}-M_B)^2}\mathrm{d}M_{D\pi}^2\, \left(M_{D\pi}^2 \sum_{l>0}\frac{W^{2l+1}}{2l+1}|\mathcal{F}_{2,l}|^2+ W\frac{(M_B^2-M^2_{D\pi})^2}{M_B^2}|f_0|^2\right)~,\\
        \Im \Pi_{A}^T\Big|_{D\pi} &= \frac{1}{192\pi^3}\frac{M_B^4}{q^2} \int_{(M_D + m_\pi)^2}^{(\sqrt{q^2}-M_B)^2}\mathrm{d}M_{D\pi}^2\, \left(\frac{M_{D\pi}^2}{\lambda_B} \sum_{l>0}\frac{W^{2l+1}}{2l+1}\left(\frac{|\mathcal{F}_{1,l}|^2}{q^2} + \frac{l+1}{l}|f_l|^2\right) + W\lambda_B\frac{|f_+|^2}{q^2 M_B^2}\right)~,\\
        \Im \Pi_{V}^T\Big|_{D\pi} &= \frac{1}{192\pi^3}\frac{M_B^4}{q^2} \int_{(M_D + m_\pi)^2}^{(\sqrt{q^2}-M_B)^2}\mathrm{d}M_{D\pi}^2\, M_{D\pi}^2 \sum_{l>0}W^{2l+1}\frac{l+1}{l(2l+1)}|g_l|^2~.\label{eq:ffunitaritybound}
    \end{split}
\end{equation}
\end{widetext}

The positivity of the squared amplitudes in the sum over states implies $\Im \Pi_{J}^{L/T}\Big|_{D\pi} \leq \Im \Pi_{J}^{L/T}$.
Inequalities for the \BDPi form factors can then be derived from this inequality by inserting Eq.~\eqref{eq:ffunitaritybound} and the perturbative expression for $\Im \Pi_{J}^{L/T}$.
These so-called unitarity bounds provide $q^2$-dependent constraints that should be incorporated in determinations of the \BDPi form factors.
Because each form factor only couples to one polarization state of the weak current in Eq.~\eqref{eq:ffunitaritybound}, the unitarity bounds are diagonal and apply only to the groups of form factors $\{f_0, \mathcal{F}_{2,l}\}$ and $\{f_+, f_l, \mathcal{F}_{1,l}\}$ and to the single form factor $g_l$, rather than more general linear combinations.
A parameterization of the $q^2$-dependence of the form factors is required to concretely specify how the bounds are imposed; we turn to this next.

\textbf{\textit{$z$-expansion and scattering constraints}  --- }
The model-independent parameterization and bounds presented in previous sections make no assumptions about the number, energies, or lineshapes of possible resonances. To render fitting the measured \BDPi decay spectra to our parameterization more tractable, it is helpful to include additional theoretical information and make some plausible assumptions.

The semileptonic $B$-decay form factors can be factorized into a part describing the short-distance weak decay and a part encoding the long-ranged final-state interactions between the hadrons~\cite{Watson:1954uc,Migdal:1956tc}:
\begin{equation}
  f_l(q^2,M_{D\pi}^2) = \hat{f}_l(q^2,M_{D\pi}^2)h_l(M_{D\pi}^2)~.\label{eq:ffexact}
\end{equation}
The weak-interaction contribution to the form factors of QCD resonances is approximately independent of $M_{D\pi}$~\cite{Shi:2020rkz}:
\begin{equation}
\hat{f}_l(q^2,M_{D\pi}^2) \approx \tilde{f}_l(q^2) + \mathcal{O}((M_R^2-M_{D\pi}^2)/M_B^2)~,\label{eq:ffaprox}
\end{equation}
Indeed, studies of $B \rightarrow \pi\pi(K)$ in the context of light cone sum rules (LCSR) \cite{Meissner:2013hya} and recent LQCD studies of the $B\rightarrow \rho$ form factors \cite{Leskovec:2022ubd} point towards the smallness of the neglected contributions.

The $q^2$-dependent function in Eq. \eqref{eq:ffaprox} can be expanded as a power series~\cite{Boyd:1995cf,Boyd:1995sq,Boyd:1997kz}
\begin{equation}
    \tilde{f}_l(q^2) = \frac{1}{\phi^{(f)}_l(q^2)B_f(q^2)}\sum_{i=0}^\infty a^{(f)}_{li} z^i,
\end{equation}
where the Blaschke product $B_f$ removes the poles of all subthreshold $B_c$ resonances for a given channel and the change of variables
\begin{equation}
    z(q^2,q^2_0) = \frac{q^2_0-q^2}{(\sqrt{q^2_+ - q^2}+\sqrt{q^2_+ - q_0^2})^2}
\end{equation}
with $q^2_+ = (M_B + M_D + m_\pi)^2$ maps the kinematically-allowed $q^2$ range onto $|z| < 1.$
With a suitable choice of outer functions $\phi^{(f)}_l$, the unitarity bounds on the $z$-expansion coefficients in Eq.~\eqref{eq:ffunitaritybound} take an especially simple form
\begin{equation}
    \sum_{i,l} |a^{(f)}_{li}|^2 < 1~,
\end{equation}
allowing for an easy integration of the form factors in a fit, including priors on the coefficients. 
Further, for $q^2_0 = 0$ GeV$^2$, $z$ ranges from $0$ to $0.06$, such that only a few terms in the expansion are needed to describe the form factors with high precision.
The truncated set of coefficients $a^{(f)}_{li}$ is then determined using fits to experimental data, fixing both shape and normalization of the form factors.

Recently discussed problems associated to lower-lying branch cut at $q^2 = (M_B+M_D)^2$ can be incorporated as outlined in Refs. \cite{Gubernari:2020eft,Gubernari:2022hxn,Blake:2022vfl,Flynn:2023qmi}. Additional details on the derivation and numerical calculation of the outer functions can be found in the supplemental material \cite{supplemental} and Ref.~\cite{rudin}.

The semileptonic-decay form factors can also be connected to the $S$-matrix via a dispersion relation~\cite{Omnes:1958hv}
\begin{equation}
\begin{split}
&\Im{\vec{f}(q^2, M_{D\pi}^2 + i\epsilon)} \\
&= T^\ast(M_{D\pi}^2 + i\epsilon)\Sigma(M_{D\pi}^2)\vec{f}(q^2, M_{D\pi}^2 + i\epsilon)~,\label{eq:disphad}
\end{split}
\end{equation}
where the form-factor $f$ is a vector in channel space,
$T$ is the $T$-matrix, and $\Sigma$ contains the relevant phase-space factors and is defined in Ref.~\cite{Omnes:1958hv}. 
The solution of Eq.~\eqref{eq:disphad} is given by the Muskhelishvili-Omnès (MO) matrix $\Omega$ \cite{Omnes:1958hv,Muskhelishvili:1953}
\begin{equation}
\begin{split}
&\vec{f}(q^2,M_{D\pi}^2) = \Omega(M_{D\pi}^2) \vec{P}(q^2,M_{D\pi}^2)~,\\
&\Im{\Omega(s + i\epsilon)} = \frac{1}{\pi}\int_{s_\mathrm{thr}}^\infty\frac{T^\ast(s')\Sigma(s')\Omega(s')}{s'-s-i\epsilon}\mathrm{d}s'~,\label{eq:omnes}
\end{split}
\end{equation}
where $\vec{P}$ are boundary functions. 
A numerical algorithm to solve the above integral equation is outlined in Ref.~\cite{Moussallam:1999aq}.

Following the same arguments as for $\tilde{f}_l$ in Eq.~\eqref{eq:ffaprox}, we can neglect the mild dependence of the boundary functions $\vec{P}$ on $M_{D\pi}$ and express them as a power-series in $z$ with the same Blaschke factors as the form factors but different outer functions.

For the S-wave contribution, we compute the MO matrix $\Omega$ from the $S$-matrix provided in Ref.~\cite{Albaladejo:2016lbb}, which was obtained using next-to-leading order UChPT interaction potentials for coupled-channel $D\pi$, $D\eta$ and $D_s K$ scattering from Ref.~\cite{Guo:2009ct,Liu:2012zya}. Consequently, the S-wave two-pole structure is treated in a parameterization-independent manner, solely relying on scattering phase-shifts. This allows us to constrain the \BDeta and $B\rightarrow D_s K \ell \nu$ decay rates from a fit of the \BDPi invariant-mass spectrum, as further discussed in the supplemental material \cite{supplemental}.
A similar procedure could be used to constrain higher-order partial-wave contributions --- albeit with additional complications due to contributions from $D^\ast\pi$ and similar channels. LQCD calculations of coupled-channel $D^\ast\pi - D\pi$ scattering amplitudes could help determine the missing ingredients.

\textbf{\textit{Experimental fits  --- }}
To test our new \BDPi form-factor description and extract the coefficients of the $z$ expansion from data, we proceed in two steps. First, we fit the measured $w$- and $\cos\theta$-dependence of the $B\rightarrow D_2^\ast \ell \nu$ differential decay width~\cite{Belle:2007uwr} and  $B^0\rightarrow D_2^{\ast-} \pi^+$ branching fraction~\cite{ParticleDataGroup:2022pth} constraining the $a_{li}^{(f)}$ with Gaussian priors centered at zero with unit width.  We employ the least-squares fitting package \verb|lsqfit| and use the augmented $\chi_{\rm aug}^2$ defined in Refs.~\cite{Lepage:2001ym,Hornbostel:2011hu} to assess the goodness of fit.  Additional numerical inputs are taken from Refs.~\cite{Miguel,FermilabLattice:2021cdg,Bigi:2016mdz,Bigi:2017njr} as discussed in the supplemental material~\cite{supplemental}. The loose constraints help the fit converge more quickly but have little impact on the final results since the magnitudes of the resulting $z$-coefficients are all of order a tenth or smaller.

\begin{figure}
    \centering
    \includegraphics[width=\linewidth]{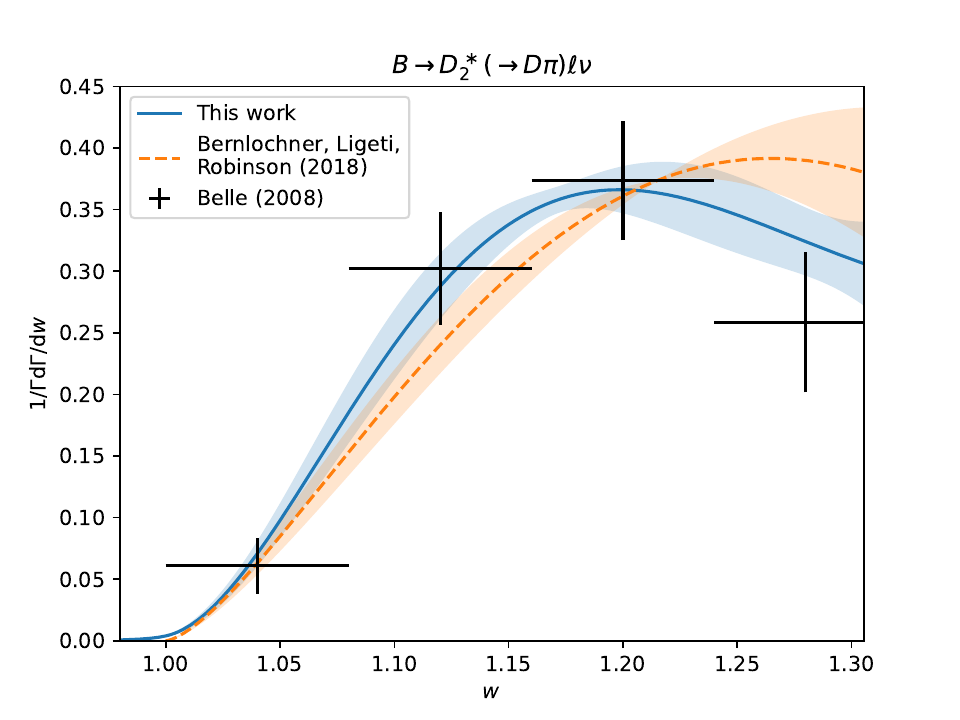}
    \caption{Normalized $B\rightarrow D_2^\ast \ell \nu$ $w$-spectrum. The black data points are from Ref.~\cite{Belle:2007uwr}. The blue solid curve with error band is our fit result, while the orange dashed curve and band are from Refs.~\cite{Bernlochner:2016bci,Bernlochner:2017jxt}. The $\chi_{\rm aug}^2/\text{dof} = 6.4/12$ and $Q = 0.9$.}
    \label{fig:D2q2}
\end{figure}

As shown in Fig. \ref{fig:D2q2}, we find a harder $D_2^\ast$ $w$-spectrum than Refs. \cite{Bernlochner:2016bci,Bernlochner:2017jxt}, {\it i.e.}  enhanced for low values of $w$ or high values of $q^2$, and also better describe the data.
Possible reasons for this difference are the use of $B\rightarrow D_1\pi$ decay data in Refs. \cite{Bernlochner:2016bci,Bernlochner:2017jxt} and the greater flexibility in the model independent approach employed here in comparison to the heavy-quark effective theory (HQET) based approach of these works.

\begin{figure}
    \centering
    \includegraphics[width=\linewidth]{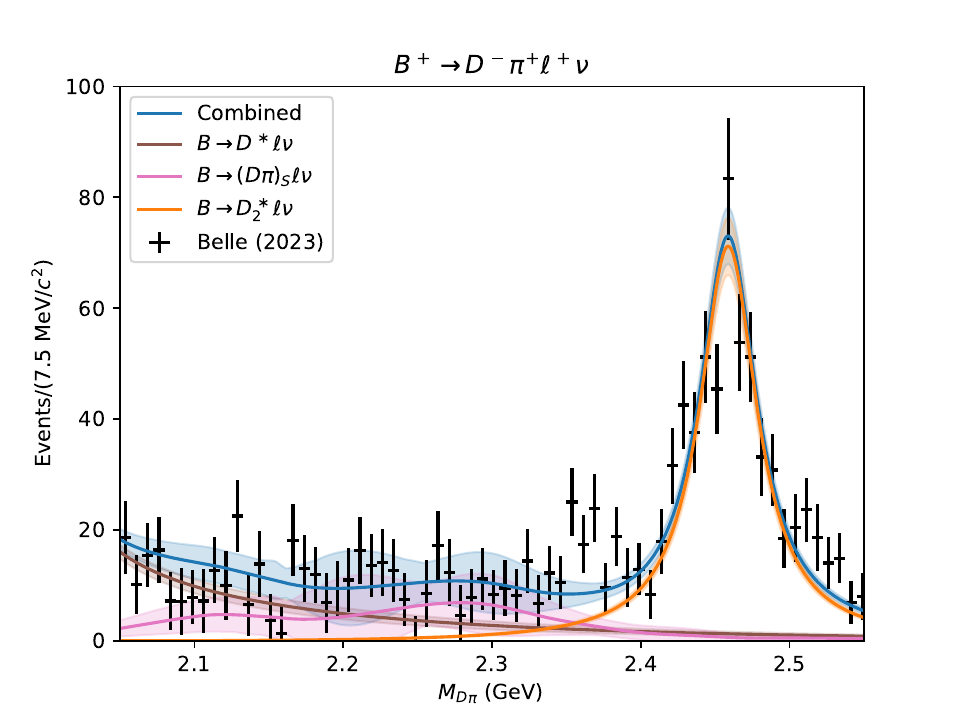}
    \caption{Fit of the measured $M_{D\pi}$-spectrum~\cite{Belle:2022yzd} using the $z$-expansion to parameterize $D_2^\ast$ and S-wave form factors. The $\chi^2_{\rm aug}/\text{dof} = 124.4/133$ and $Q = 0.69$. Only data for the more precise $B^+$ mode is shown.}
    \label{fig:massspec}
\end{figure}

Next, we fit the \BDPi $M_{D\pi}$-spectrum  measured recently by Belle~\cite{Belle:2022yzd} using the $z$-expansion coefficients from the first fit as priors to constrain the shape the $D_2^\ast$ form factors.  Following Refs.~\cite{LeYaouanc:2018zyo,LeYaouanc:2022dmc}, we parameterize the $D^\ast$ and $D^\ast_2$ lineshapes by a Breit-Wigner distribution and with Blatt-Weisskopf damping factors~\cite{Blatt:1952ije,VonHippel:1972fg}. In contrast to Ref.~\cite{LeYaouanc:2022dmc} we allow the Blatt-Weisskopf radius to be determined in the fit. As shown in Fig.~\ref{fig:massspec}, our form-factor parameterization provides a good description of the data over the entire invariant-mass range.

\begin{figure}
    \centering
    \includegraphics[width=\linewidth]{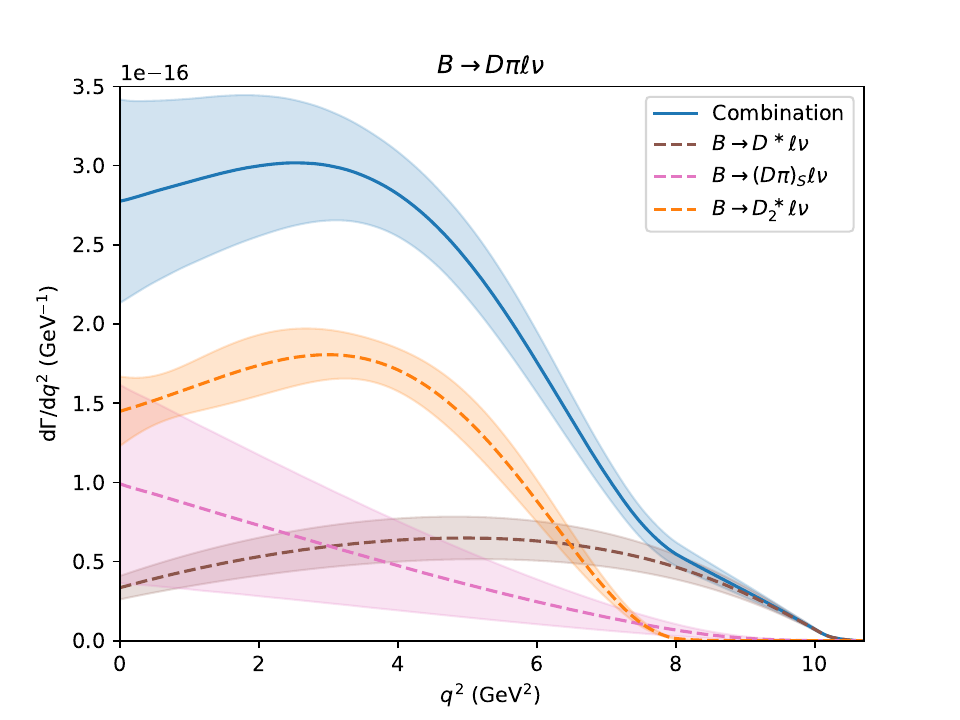}
    \caption{Predicted partial-wave decomposition of the \BDPi $q^2$-spectrum (dashed and dotted curves with error bands) and their total (solid curve with error band) from the fit in  Fig.~\ref{fig:massspec}.}
    \label{fig:q2spec}
\end{figure}

Our fit to the \BDPi invariant-mass spectrum can be used to make predictions for related quantities.  Figure~\ref{fig:q2spec} shows the predicted partial-wave contributions to the $q^2$-spectrum. After integrating over the momentum transfer, we obtain for the D-wave channel $\mathrm{Br}(B\rightarrow D_2^\ast(\rightarrow D \pi^\pm) \ell \nu) = (1.90\pm 0.11)\times 10^{-3}$, which is larger than Belle's determination~ in Ref.~\cite{Belle:2022yzd}. This is because the smooth falling function employed by Belle to describe the seemingly nonresonant contributions overlaps with the $D_2^\ast$ resonance, whereas in our description, the S-wave and $D^\ast$ components are negligible near the resonance. For the S-wave contribution, we obtain $\mathrm{Br}(B \rightarrow (D \pi)_S \ell \nu) = (1.03 \pm 0.27) \times 10^{-3}$, which agrees with the arguments made in Ref.~\cite{LeYaouanc:2022dmc} but is smaller than the branching fraction usually assigned  in experimental analyses.
Finally, the $P$-wave contribution in $B^+ \rightarrow D^- \pi^+ \ell^+ \nu_\ell$ decays, to which on-shell $D^\ast$ decays can not contribute, amounts to a branching ratio of $(9.2\pm 0.9)\times 10^{-4}$ for $M_{D\pi}\leq 2.5$\,GeV.

\textbf{\textit{Implications and outlook --- }}
We present the first model-independent description of \BDPi decays based on unitarity and analyticity of the relevant form factors and the factorization of final-state interactions. This constitutes the first generalization of the BGL parameterization to multi-hadron final states and provides the first step towards a model-independent study of semileptonic $B$-meson decays into higher resonances and non-resonant final states. 
Our framework does not include any assumptions about lineshapes of resonances and is extendable to other decay processes with charmed mesons in the final state such as $B\rightarrow D^\ast \pi \ell \nu$ or $B_s\rightarrow D K \ell \nu$. Further, it is also valid for final states with more than two hadrons, and can be combined with other known $b\rightarrow c$ form factors in a global fit to obtain constraints on less well-known form factors (see e.g Ref.~\cite{Cohen:2019zev}).
By replacing the $D$ meson by a pion, the unitarity bounds can be applied to $B\rightarrow \pi \pi \ell \nu$ decays, including the phenomenologically interesting $B\rightarrow \rho \ell \nu$ channel, which is the target of first LQCD calculations beyond the narrow-width limit \cite{Leskovec:2022ubd}, as well as non-resonant backgrounds, which constitute the dominant systematic uncertainty~\cite{Belle-II:2022fsw}.

Taking into account recent theoretical considerations and measurements of \BDPi decays by Belle, we provide precise predictions for semileptonic decays
into the broad two-pole structure in the S-wave and determine the form-factor parameters for $B \rightarrow D_2^\ast \ell \nu$ decays from data. 
This marks the first time in which a three-component hypothesis consisting of S-wave contributions, $D^\ast$ virtual contributions and $D_2^\ast$ contributions, is compared to the measured $M_{D\pi}$-spectrum. Previous works either do not include all three components simultaneously \cite{Belle:2007uwr,BaBar:2007ddh,Bernlochner:2016bci} or do not compare to the measured $M_{D\pi}$-spectra \cite{LeYaouanc:2022dmc}.
We demonstrate, in contrast to existing literature, that our treatment of the S-wave is compatible with the $M_{D \pi}$-spectrum measured by Belle and, since it is the clearly favored description of $B^+ \rightarrow D^- \pi^+\pi^+$ decays \cite{Du:2020pui}, should replace models that assume a single, broad S-wave state, the $D_0^\ast(2300)$.
While more careful studies need to be conducted, the change in the shape for $B\rightarrow D_2^\ast \ell \nu$ decays, as well as the inclusion of the virtual $D^\ast$ contribution lead to an overall harder $q^2$-spectrum, potentially resolving some of the discrepancies seen in inclusive analyses at high $q^2$ \cite{Belle:2021eni,Belle:2021idw,Belle-II:2023aih}.

The coupled-channel nature of the S-wave contribution enables us to obtain predictions for \BDeta and $B \rightarrow D_s K \ell \nu$ decays purely based on measurements of \BDPi decays.
For the $D\eta$ S-wave contribution we obtain $\mathrm{Br}(B \rightarrow (D \eta)_S \ell \nu) = (1.9 \pm 1.7) \times 10^{-5}$, two orders of magnitude too small to constitute a sizeable portion of the semileptonic gap.
Since heavy-quark spin symmetry relates the S-wave scattering matrix to the $D^\ast\pi-D^\ast\eta-D_s^\ast K$ $J^P = 1^+$ S-wave scattering matrix, the same conclusion holds for the $B \rightarrow (D^\ast \eta)_S \ell \nu)$ channel. Consequently, both approaches utilized by the Belle and Belle II Collaborations in recent measurements to fill the semileptonic gap in terms of $B\rightarrow D^{(\ast)} \eta \ell \nu$ decays, either via a broad S-wave resonance or equidistributed in phase space, are ruled out. While our analysis does not provide alternative candidates to fill the gap, the harder $q^2$-spectrum of $B\rightarrow D \pi \ell \nu$ decays obtained here shift the gap to lower values of $q^2$, thus opening up the possibility of heavier states accounting for it.

Additional theoretical work, such as a more precise determination of the scattering potentials along the lines of Ref.~\cite{Liu:2012zya}, LQCD determinations of the form factors and LCSR computations of the S-wave form factors \cite{Gubernari:2023rfu} would greatly improve the results presented in this letter.

Future experimental measurements of the $q^2$- and $\cos\theta_l$-spectra of  $B \rightarrow D_2^\ast \ell \nu$ decays by Belle II with the already available data set, as well as updated angular analyses of $B^0 \rightarrow D^0 \pi^- \pi^+$ and $B^0 \rightarrow D^0 \pi^- K^+$ decays by LHCb, would improve the precision of the form factors presented in this letter. In the long term, a full partial-wave analysis of \BDPi decays is required to ultimately determine the exact composition of the $D\pi$ spectrum in semileptonic decays. Additionally, the final state interactions between $D$ mesons and light hadrons can be tested by measuring femtoscopic correlation functions at the ALICE experiment \cite{ALICE:2022enj}. This result could provide a direct, orthogonal test of the S-wave two-pole structure in heavy ion collisions \cite{Albaladejo:2023pzq}.

\section{Acknowledgments}
We thank Dmitri Liventsev and Frank Meier for discussions and providing us with the data from Refs. \cite{Belle:2007uwr} and \cite{Belle:2022yzd}, respectively, and Miguel Albaladejo for providing us with code to reproduce the results of Ref.~\cite{Albaladejo:2016lbb}.
We also thank Feng-Kun Guo, Christoph Hanhart, Andreas Kronfeld, Bastian Kubis, Hank Lamm, and Luka Leskovec for helpful discussions on the subject matter and feedback on the draft. RvT thanks Fermilab for hospitality during the completion of this work.
Fermilab is operated by Fermi Research Alliance, LLC under contract number DE-AC02-07CH11359 with the United States Department of Energy. FH acknowledges support by the Alexander von Humboldt foundation. RvT acknowledges support by the Natural Sciences and Engineering Research Council of Canada. RvT dedicates this paper to her grandfather, Abraham Petrus Louw Tonkin, who sadly passed away during the completion of this manuscript. We'll miss your tall tales.

\bibliography{refs}

\input{suppl}

\end{document}

%% file: suppl.tex
\newpage

\onecolumngrid
\newpage
\appendix

\section*{Supplemental material}
\subsection{Angular momentum vectors}
By demanding Eq.~\eqref{eq:Lvec} to be fulfilled, the vectors $L^{(l)}$  with $l>0$ in the form-factor decomposition Eq.~\eqref{eq:ffs1} can be systematically and uniquely constructed.
Orthogonality to $p^\mu_{D\pi}$ is enforced by the applying the projectors
\begin{equation}
\begin{split}
    P^{\mu\nu} &= \left( -g^{\mu\mu} + \frac{p_{D\pi}^\mu p_{D\pi}^\nu}{M_{D_\pi}^2}\right)~
\end{split}
\end{equation}
to $q^\mu$ and the other linearly independent momentum combination
\begin{equation}
        \delta^\mu = \frac{1}{2M_B}(p_D - p_\pi)^\mu~.
\end{equation}
It is easy to show that
\begin{equation}
    \begin{split}
    q_\mu P^{\mu\nu}\delta_\nu &= M_B W \cos\theta~,\\
    q_\mu P^{\mu\nu} q_\nu &= \frac{M_B^2|\vec{p}_{D\pi}|^2}{M_{D\pi}^2}~,\\
    \delta_\mu P^{\mu\nu}\delta_\nu &= \frac{|\vec{p}_{D}|^2}{M_B^2}~,
    \end{split}
\end{equation}
where the three-momenta are defined in the $B$- and $D\pi$-restframes, respectively.
Consequently, for a given partial wave $l$, Eq.~\eqref{eq:Lvec} requires $L^{(l)}_{\mu}$ to contain $l$ powers of $\delta^\mu$ and $l-1$ powers of $q_\mu$ contracted with projectors $P^{\mu\nu}$. The constraint $L^{(l)}_{\mu} q^\mu \propto P_l(\cos\theta)$ then leads to a unique solution for $L^{(l)}_{\mu}$ that for example is given for $l \in \{1,2,3\}$ by 
\begin{equation}
\begin{split}
    L^{(1),\mu} &= P^{\mu\nu}\delta_\nu,\\
    L^{(2),\mu} &= \frac{3}{2 M_B}\left[P^{\mu\nu}\delta_\nu(\delta\cdot P \cdot q) - \frac{1}{3} P^{\mu\nu}q_\nu(\delta\cdot P \cdot \delta) \right],\\
    L^{(3),\mu} &= \frac{5}{2 M_B^2}\Bigg[P^{\mu\nu}\delta_\nu\left((\delta\cdot P \cdot q)^2-\frac{1}{5}(q\cdot P \cdot q)(\delta\cdot P \cdot \delta)\right) - \frac{2}{5} P^{\mu\nu}q_\nu(q\cdot P \cdot \delta)(\delta\cdot P \cdot \delta) \Bigg]~.
\end{split}
\end{equation}

\subsection{Fivefold differential decay rate}
The fivefold differential decay rate for \BDPi decays is given by
\begin{equation}
\frac{\mathrm{d}\Gamma_{B\rightarrow D\pi\ell\nu}}{\mathrm{d}M_{D\pi}^2\mathrm{d}q^2\mathrm{d}\cos\theta\mathrm{d}\cos\theta_l\mathrm{d}\chi} = \frac{G_F^2|V_{cb}|^2}{M_B} \frac{1}{(4\pi)^6}\left(1-\frac{m_l^2}{q^2}\right)W\sum_{a,b}\left|\mathcal{M}_{ab}\right|^2~,
\end{equation}
where $a,b\,\in \{0,1,2\}$ label the partial waves and $G_F$ is Fermi's constant.
The matrix element squared is given by
\begin{equation}
    \left|\mathcal{M}_{ab}\right|^2 = \left\langle D(p_D) \pi(p_\pi)|V_\mu-A_\mu|B(p_B)\right\rangle \left\langle D(p_D) \pi(p_\pi)|V_\nu-A_\nu|B(p_B)\right\rangle L^{\mu\nu}(p_l,p_\nu)
\end{equation}
where
\begin{equation}
    L^{\mu\nu}(p_l,p_\nu) = \mathrm{Tr}\left(\gamma^\mu P_L (\slashed{p}_l - m_l) \gamma^\nu P_L \slashed{p}_\nu\right)
\end{equation}
is the leptonic tensor and $P_L = (\mathbb{1}-\gamma_5)/2$.

Inserting Eq.~\eqref{eq:ffs1}, we obtain
\begin{equation}
\begin{split}
\left|\mathcal{M}_{ab}\right|^2 
&= M_B^2 M_{D\pi}^2(q^2-m_l^2)W^{a+b} \Bigg\{ \left[\frac{\tilde{\mathcal{F}}_{1,a}\tilde{\mathcal{F}}^\ast_{1,b}}{\lambda(M_B^2,M_{D\pi}^2,q^2)q^2} + \frac{m_l^2}{q^4}\mathcal{F}_{2,a}\mathcal{F}^\ast_{2,b} - \frac{f_a f^\ast_b}{\lambda(M_B^2,M_{D\pi}^2,q^2)} - g_a g^\ast_b\right]P_a^0 P_b^0 \\ &- \left(1-\frac{m_l^2}{q^2}\right)\frac{\tilde{\mathcal{F}}_{1,a}\tilde{\mathcal{F}}^\ast_{1,b}}{\lambda(M_B^2,M_{D\pi}^2,q^2)q^2}\cos^2\theta_l P_a^0 P_b^0\\ &+\left(\frac{f_a f^\ast_b}{\lambda(M_B^2,M_{D\pi}^2,q^2)} + g_a g^\ast_b\right)\left[P_{a-1}^0P_{b-1}^0 + \frac{P_{a-1}^1P_{b-1}^1}{ab}\right] - \left(1-\frac{m_l^2}{q^2}\right)g_a g^\ast_b\frac{P_a^1P_b^1}{ab}(1-\cos^2\theta_l)\\ &- \left(1-\frac{m_l^2}{q^2}\right)\left(\frac{f_a f^\ast_b}{\lambda(M_B^2,M_{D\pi}^2,q^2)} - g_a g^\ast_b\right)\frac{P_a^1P_b^1}{ab}\cos^2\chi(1-\cos^2\theta_l)\\
&-\frac{f_a g^\ast_b + g_a f^\ast_b}{\sqrt{\lambda(M_B^2,M_{D\pi}^2,q^2)}}\left[P_{a-1}^0P_{b-1}^0 + \frac{P_{a-1}^1P_{b-1}^1}{ab}\right]\cos\theta_l + \left(\frac{f_a g^\ast_b + g_a f^\ast_b}{\sqrt{\lambda(M_B^2,M_{D\pi}^2,q^2)}} + \frac{m_l^2}{q^4}\frac{\tilde{\mathcal{F}}_{1,a}\mathcal{F}^\ast_{2,b}+\mathcal{F}_{2,a}\tilde{\mathcal{F}}^\ast_{1,b}}{\sqrt{\lambda(M_B^2,M_{D\pi}^2,q^2)}}\right)\cos\theta_l P_a^0 P_b^0\\
&+i\left(1-\frac{m_l^2}{q^2}\right)\frac{g_a f_b^\ast - f_a g_b^\ast}{\sqrt{\lambda(M_B^2,M_{D\pi}^2,q^2)}}\frac{P_a^1P_b^1}{ab}(1-\cos^2\theta_l)\cos\chi\sin\chi\\
&+\left(1-\frac{m_l^2}{q^2}\right)\left[\frac{\tilde{\mathcal{F}}_{1,a}f_b^\ast}{\sqrt{q^2}\lambda(M_B^2,M_{D\pi}^2,q^2)}\frac{P_a^0 P_b^1}{b}+\frac{f_a\tilde{\mathcal{F}}^\ast_{1,b}}{\sqrt{q^2}\lambda(M_B^2,M_{D\pi}^2,q^2)}\frac{P_a^1 P_b^0}{a}\right]\cos\theta_l\sin\theta_l\cos\chi\\
&+i\left(1-\frac{m_l^2}{q^2}\right)\left[\frac{\tilde{\mathcal{F}}_{1,a}g_b^\ast}{\sqrt{q^2}\sqrt{\lambda(M_B^2,M_{D\pi}^2,q^2)}}\frac{P_a^0 P_b^1}{b}-\frac{g_a\tilde{\mathcal{F}}^\ast_{1,b}}{\sqrt{q^2}\sqrt{\lambda(M_B^2,M_{D\pi}^2,q^2)}}\frac{P_a^1 P_b^0}{a}\right]\cos\theta_l\sin\theta_l\sin\chi\\
&-\left[\left(\tilde{\mathcal{F}}_{1,a}g_b^\ast + \frac{m_l^2}{q^2}
\mathcal{F}_{2,a}f_b^\ast\right)\frac{P_a^0 P_b^1}{b} + \left(g_a\tilde{\mathcal{F}}^\ast_{1,b} + \frac{m_l^2}{q^2}f_a\mathcal{F}^\ast_{2,b}\right)\frac{P_a^1 P_b^0}{a}\right]\frac{\cos\chi\sin\theta_l}{\sqrt{q^2}\sqrt{\lambda(M_B^2,M_{D\pi}^2,q^2)}}\\
&-i\left[\left(\tilde{\mathcal{F}}_{1,a}f_b^\ast + \frac{m_l^2}{q^2}\mathcal{F}_{2,a}g_b^\ast\right)\frac{P_a^0 P_b^1}{b} - \left(f_a\tilde{\mathcal{F}}^\ast_{1,b} + \frac{m_l^2}{q^2}g_a\mathcal{F}^\ast_{2,b}\right)\frac{P_a^1 P_b^0}{a}\right]\frac{\sin\chi\sin\theta_l}{\sqrt{q^2}\lambda(M_B^2,M_{D\pi}^2,q^2)}\Bigg\}~,
\end{split}
\end{equation}
where we have introduced $\tilde{\mathcal{F}}_{1,i} = (M_B^2 - M_{D\pi}^2)\mathcal{F}_{1,i}$ for brevity. The $P_a^i$ are the associated Legendre polynomials with arguments of $\cos\theta$. For the S-wave contribution, i.e. $a = 0$ or $b = 0$, we have $g_a = f_a = 0$, thus all terms proportional to $1/a$ or $1/b$ do not appear, as well as $\tilde{\mathcal{F}}_{1,0} = \lambda_B/M_{D\pi} f_+$ and $\tilde{\mathcal{F}}_{2,0} = (M_B^2-M_{D\pi}^2)/M_{D\pi}f_0$.
This fully-general expression is valid for massive leptons and incorporates all interference terms between different partial waves.

\subsection{Unitarity bounds and outer functions}
Inserting Eq.~\eqref{eq:ffunitaritybound} into Eq.~\eqref{eq:disp} at $Q^2 = 0$ results in unitarity bounds that all can be cast into the form
\begin{equation}
    \chi \geq \frac{1}{\pi}\int_0^\infty\mathrm{d}q^2 \frac{M_B^4}{192\pi^3} \frac{C^{(l)}}{(2l+1)(q^2)^a}\int_{(M_D+m_\pi)^2}^{(M_B-\sqrt{q^2})^2}\mathrm{d}M_{D\pi}\frac{(M^2_{D\pi})^b (M_B^2-M^2_{D\pi})^c}{ \lambda_B^d} W^{2l+1}|f_l(q^2,M^2_{D\pi})|^2~.\label{eq:suppl:ffbounds}
\end{equation}
The individual constants and powers for all form factors can be found in Table~\ref{tbl:coeffs}.
\begin{table}
\caption{\label{tbl:coeffs} Coefficients and powers in Eq.~\eqref{eq:suppl:ffbounds} for the individual form factors.}
\begin{tabular}{ c || c | c | c | c | c | c}
 & $g_l$ & $f_l$ & $\mathcal{F}_{1,l}$ & $\mathcal{F}_{2,l}$ & $f_+$ & $f_0$\\
\hline
a & 4 & 4 & 5 & 4 & 5 & 4 \\
b & 1 & 1 & 1 & 1 & 0 & 0 \\
c & 0 & 0 & 2 & 0 & 0 & 2 \\
d & 0 & 1 & 1 & 0 & -1 & 0 \\
$\chi$ & $\chi_{(V)}^T$ & $\chi_{(A)}^T$ & $\chi_{(A)}^T$ & $\chi_{(A)}^L$ & $\chi_{(A)}^T$ & $\chi_{(A)}^L$ \\
$C^{(l)}$ & $\frac{l+1}{l}$ & $\frac{l+1}{l}$ & 1 & 3 & $\frac{1}{M_B^2}$ & $\frac{3}{M_B^2}$
\end{tabular}
\end{table}

Inserting Eq.~\eqref{eq:ffaprox} into Eq.~\eqref{eq:suppl:ffbounds} and dividing by $\chi$, we obtain
\begin{equation}
    1 \geq \frac{1}{\pi}\int_0^\infty\mathrm{d}q^2 \frac{M_B^4}{192\pi^3\chi}\frac{C^{(l)}}{(2l+1)(q^2)^a} \mathcal{I}^{(l)}_{(b,c,d)}(q^2) |\tilde{f}_l(q^2)|^2~,\label{eq:suppl:ffbounds2}
\end{equation}
where
\begin{equation}
\mathcal{I}^{(l)}_{(b,c,d)}(q^2) = \int_{(M_D+m_\pi)^2}^{(M_B-\sqrt{q^2})^2}\mathrm{d}M_{D\pi}\frac{(M^2_{D\pi})^b (M_B^2-M^2_{D\pi})^c}{ \lambda_B^d} W^{2l+1}|h_l(M^2_{D\pi})|^2~.
\end{equation}

Following Ref.~\cite{Boyd:1995cf,Boyd:1995sq,Boyd:1997kz} we write Eq.~\eqref{eq:suppl:ffbounds2} in terms of $z$ and introduce Blaschke products $B_f$ to take into account sub-threshold $B_c$-meson poles:
\begin{equation}
1 \geq \frac{1}{2\pi i}\oint_{|z|=1}\frac{\mathrm{d}z}{z} \left| B_f(z) \Phi_l^{(f)}(z) \tilde{f}_l(z)\right|^2~. \label{eq:Blaschke}
\end{equation}
The Blaschke products fulfill $|B_f| = 1$ for $|z| = 1$ and are given by
\begin{equation}
    B_f(z) = \prod_{p}z(q^2,M^2_p)~,
\end{equation}
where the product over poles $p$ runs over the $B_c$-meson poles of the relevant spin and parity.
The outer functions are defined on the unit circle by
\begin{equation}
    \left|\Phi_l^{(f)}(z)\right|^2\Bigg|_{|z| = 1} = \left(\left|\frac{\mathrm{d}z}{\mathrm{d}q^2}\right|\right)^{-1}\frac{M_B^4}{192\pi^3\chi}\frac{C^{(l)}}{(2l+1)(q^2)^a} \mathcal{I}^{(l)}_{(b,c,d)}(q^2)~.
\end{equation}
To continue the outer functions to the full unit disk, we need to introduce Blaschke products to remove zeroes and poles in $q^2$ that are mapped to $|z| < 1$.
Thus, we need to replace
\begin{equation}
    \frac{1}{(q^2)^a} \rightarrow \left(-\frac{z(q^2,0)}{q^2}\right)^a
\end{equation}
and
\begin{equation}
    \mathcal{I}^{(l)}_{(b,c,d)}(q^2) = (q^2_- - q^2)^{2 + 2l - d}\tilde{\mathcal{I}}^{(l)}_{(b,c,d)}(q^2) \rightarrow \left(\frac{(q^2_- - q^2)}{z(q^2,q^2_-)}\right)^{2 + 2l - d}\tilde{\mathcal{I}}^{(l)}_{(b,c,d)}(q^2)~,
\end{equation}
where $q^2_- = (M_B - M_D - m_\pi)^2$ is the endpoint of the semileptonic $q^2$ spectrum. 
This replacement does not modify the integrand anywhere along the integration contour in Eq.~\eqref{eq:Blaschke} and the unitarity bounds therefore still apply identically after this replacement.
The function $\tilde{\mathcal{I}}^{(l)}_{(b,c,d)}(q^2)$ is non-vanishing for any $q^2$, but is complex outside physical regions.

\subsection{Fit of experimental data}
To fit to the available data from the Belle experiment \cite{Belle:2007uwr,Belle:2022yzd} we need to parametrize the lineshapes of the $D^\ast$ and $D_2^\ast$. To this end, we write the $M_{D\pi}$-dependent part of form factors such as $g^{(l)}$ in the decomposition of Eq.~\eqref{eq:ffexact} as
\begin{equation}
    h_l(M_{D\pi}) = \frac{1}{(M_{D\pi}^2-M_{R,l}^2) + i M_{R,l} \Gamma_R(M_{D\pi}^2)}X^{(l)}(|\vec{p}_D|r_\mathrm{BW},|\vec{p}_{D,0}|r_\mathrm{BW})~.
\end{equation}
Here, $M_{R,l}$ are the respective nominal resonance masses, $\Gamma_R$ the energy-dependent {\color{blue} total} width and the $X^{(l)}$ are Blatt-Weisskopf damping factors. We take the masses, widths and branching ratios for the $D^\ast$ and $D_2^\ast$ from Ref.~\cite{ParticleDataGroup:2022pth} and determine $c_l$ from them. For the $D^\ast$ width, we take into account the $D\pi$ and $D\gamma$ decay channels and for the $D_2^\ast$, we assume the total width is composed by the $D\pi$ and $D^\ast \pi$ channels.
The damping factors are given by
\begin{equation}
    X^{(1)}(z,z_0) = \sqrt{\frac{1+z_0^2}{1+z^2}}~,\qquad
    X^{(2)}(z,z_0) = \sqrt{\frac{9+3z_0^2+z_0^4}{9+3z^2+z^4}}~.
\end{equation}
The $D$-meson three-momentum is evaluated in the resonance center-of-mass frame, $|\vec{p}_{D,0}|$ denotes the $D$-meson three-momentum for $M_{D\pi} = M_R$. The Blatt-Weisskopf radius $r_\mathrm{BW}$ is treated as a parameter in the fit with a prior of $4\pm 1$ GeV$^{-1}$, where the central value corresponds to the value chosen by the LHCb experiment for their amplitude analysis of $B^- \rightarrow D^+ \pi^- \pi^-$ decays \cite{LHCb:2016lxy}. No additional phase-space factors need to be included here, as all partial-wave specific factors are included in the angular momentum vectors $L^{(l)}$ through the powers of $W$ in Eq.~\eqref{eq:Lvec}.

Uncertainties in the scattering phase shifts from Ref.~\cite{Albaladejo:2016lbb} are obtained by computing our results using 150 samples provided by the authors of Ref.~\cite{Liu:2012zya} that describe the combined $1\sigma$ confidence contour of the parameters fitted in that work.  We compute uncertainties from the variation of our results across these samples, and combine these uncertainties in quadrature with the other uncertainties appearing in our analysis.

To evaluate the integral Eq.~\eqref{eq:omnes}, we need to continue the $T$-matrix from the region described by the phase shifts and inelasticities from Ref.~\cite{Albaladejo:2016lbb} to $\infty$. We follow Ref.~\cite{Albaladejo:2016lbb} with one difference: we do not demand that
\begin{equation}
    \sum^3_{i=1} \delta_i(\infty) = 3 \pi~,
\end{equation}
but rather continue the $D\pi$ phase-shift to $\pi$ and the other two to $0$.
This avoids additional resonances within the semileptonic region when the $D\pi$ phase-shift takes on values of $(2n+1)/2\pi$ and minimizes the influence of the function chosen to continue the phase shifts. This treatment is equivalent, within uncertainties, to continuing $D\pi$ phase-shift to $3\pi$, as in Ref.~\cite{Albaladejo:2016lbb}, but slow enough, to push the additional unphysical resonances to energies much larger than the semileptonic region.

In the following, we describe our fit procedure step by step.
First, we subtract the contribution of $B_c$ meson states for which the leptonic decay constants are known from the susceptibilities to strengthen the unitarity bounds following Refs.~\cite{Bigi:2016mdz,Bigi:2017njr} and truncate the $z$-expansion of the form factors at linear order in $z$. All $z$-expansion coefficients are sampled from gaussian priors centered around zero with unit width to take into account the overall scale set by the unitarity bounds.

To obtain the P-Wave form factors, we integrate the five-fold differential decay rate over $M_{D\pi}$ and match the four-fold differential decay rate to the one obtained from the LQCD determination of the $B\rightarrow D^\ast \ell \nu$ form factors by the Fermilab/MILC collaboration \cite{FermilabLattice:2021cdg}:
\begin{equation}
    \int\mathrm{d}M_{D\pi}^2 \frac{\mathrm{d}^5 \Gamma}{\mathrm{d}M_{D\pi}^2\mathrm{d}q^2\mathrm{d}\cos\theta\mathrm{d}\cos\theta_l\mathrm{d}\chi} = \mathrm{Br}(D^\ast \rightarrow D \pi)\frac{\mathrm{d}^4 \Gamma}{\mathrm{d}q^2\mathrm{d}\cos\theta\mathrm{d}\cos\theta_l\mathrm{d}\chi}\Bigg|_\text{FNAL/MILC}~.
\end{equation}

Next, we obtain the D-Wave form factor parameters by performing a simultaneous fit to the $w$- and $|\cos\theta|$- spectra from Ref.~\cite{Belle:2007uwr}, as well as the world averages for the nonleptonic rates for $B^0 \rightarrow D_2^{\ast -} \pi^+$ and $B^0 \rightarrow D_2^{\ast -} K^+$ \cite{ParticleDataGroup:2022pth}.
Following Refs.~\cite{Bernlochner:2016bci,Bernlochner:2017jxt,LeYaouanc:2022dmc}, we assume naive factorization and obtain the double-differential decay rate from Eq.~\eqref{eq:ffs1}:
\begin{equation}
    \frac{\mathrm{d}\Gamma_{B\rightarrow D \pi P}}{\mathrm{d}M_{D\pi}^2\mathrm{d\cos\theta}} = \frac{G_F^2 |V_{cb}|^2 M_B}{(4\pi)^3}\frac{f_P^2 |V_{ux}|^2 |a_\text{eff}|^2}{4}M_{D\pi}^2\sum_{a,b} W^{a+b+1}\mathcal{F}_{2,a}\mathcal{F}^\ast_{2,b}P_a^0 P_b^0\Bigg|_{q^2 = m_P^2}~.
\end{equation}
Here $P = \pi, K$, $x = d, s$ and we take $a_\text{eff} = 0.93 \pm 0.07$~\cite{LeYaouanc:2022dmc}.

We then perform a fit to the $M_{D\pi}$-spectra of Ref.~\cite{Belle:2022yzd} with the S-, P- and D-wave form factor parameters entering the fit with priors set to the previously extracted values and covariances, masses and widths constrained by their world averages and $r_\mathrm{BW}$ treated as described above. 
Finally, we test if the unitarity bounds are satisfied.

Both charge modes are fitted simultaneously, unlike in experimental fits to the mass spectrum, with the sole difference being the exact values of the involved masses. Since $D^{\ast 0} \rightarrow D^+ \pi^-$ decays are not allowed on-shell, this leads to slightly different shapes in the $M_{D\pi}$ spectrum at low masses. As a cross-check, we also considered the $M_{D\pi}$-spectra of Ref.~\cite{Belle:2007uwr}, finding good agreement and compatible results with the main fit.

Similarly to the normalized $q^2$ spectrum in $B\rightarrow D_2^\ast \ell \nu$ decays shown in Fig.~\ref{fig:D2q2}, the normalized $q^2$ spectrum we obtain for the S-wave contribution is slightly enhanced at high $q^2$ compared to the $D_0^\ast$ contribution obtained in Refs.~\cite{Bernlochner:2016bci,Bernlochner:2017jxt}. However, in contrast to the $D_2^\ast$ contribution, the resulting spectra are better compatible with each other due to the larger uncertainties.
\begin{figure}
    \centering
    \includegraphics[width=0.5\linewidth]{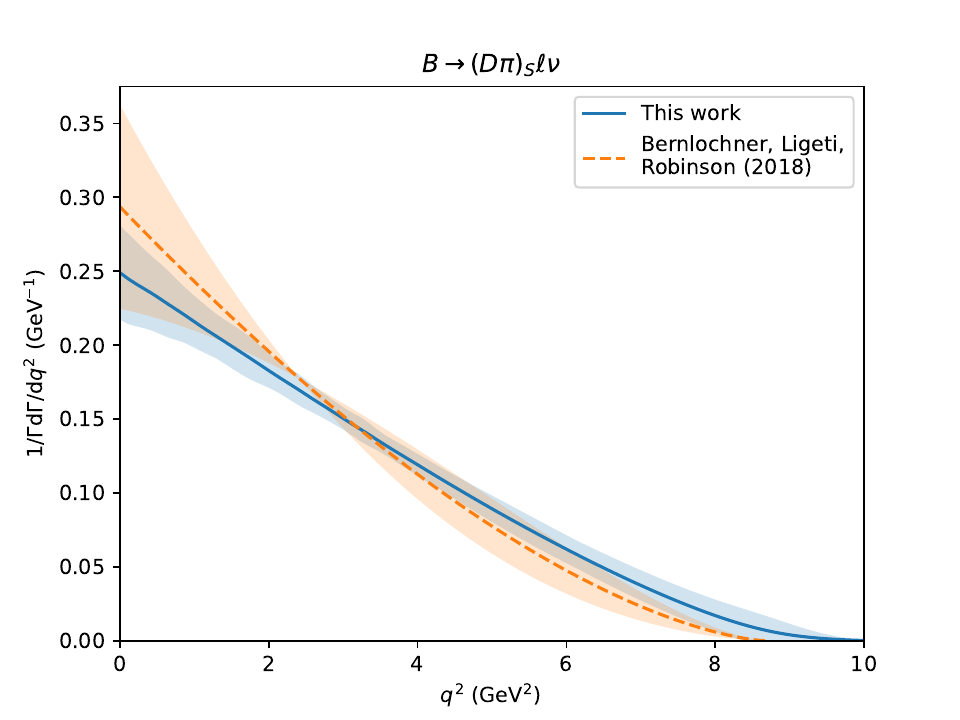}
    \caption{Normalized $B\rightarrow (D \pi)_S \ell \nu$ $q^2$-spectrum. The blue solid curve with error band is our fit result, while the orange dashed curve and band correspond to the $B\rightarrow D_0^\ast \ell \nu$ spectrum from Refs.~\cite{Bernlochner:2016bci,Bernlochner:2017jxt}.}
    \label{fig:D0q2}
\end{figure}
As the S-wave fit constrains all three components of $\vec{P}$, defined in Eq.~\eqref{eq:omnes}, we automatically obtain the full three-channel form-factor $\vec{f}$. After integration over the invariant mass of the hadronic system and $q^2$, we obtain the decay rates for semileptonic decays with S-wave $D_s K$ and $D\eta$ systems.